\documentclass[pra,twocolumn,superscriptaddress]{revtex4} 
\usepackage{epsfig}

\usepackage{graphicx}

\usepackage{color,colordvi,graphicx,pstcol}

\newcommand{\myem}[1]{({\sc\large #1})}

\newcommand{\beq}{\begin{equation}}
\newcommand{\eeq}{\end{equation}}

\newcommand{\hide}[1]{}

\newcommand{\eq}[1]{Eq.\,(\ref{#1})}

\newcommand{\fig}[1]{Fig.\,\ref{#1}}

\begin{document}
\title{Rotation of atoms in a two dimensional lattice with a harmonic trap}
\author{T. Wang}
\affiliation{Department of Physics, University of Connecticut,
Storrs, CT 06269}

\author{S. F. Yelin}
\affiliation{Department of Physics, University of Connecticut,
Storrs, CT 06269} \affiliation{ITAMP, Harvard-Smithsonian Center
for Astrophysics, Cambridge, MA 02138}

\date{\today}
\begin{abstract}
Rotation of atoms in a lattice is studied using a Hubbard model.
It is found that the atoms are still contained in the trap even
when the rotation frequency is larger than the trapping frequency.
This is very different from the behavior in continuum. Bragg
scattering and coupling between angular and radial motion are
believed to make this stability possible. In this regime, density
depletion at the center of the trap can be developed for spin
polarized fermions.
\end{abstract}
\maketitle

Recent spectacular progress in manipulating neutral atoms has
opened the way to the simulations of complex quantum systems of
condensed mater physics, such as high-Tc superconductors, by means
of atomic systems with perfectly controllable physical parameters.
The systems studied include Bose-Einstein condensations (BEC) for
both atoms and molecules, paired states for fermions (BCS), the
crossover between BEC and BCS
~\cite{Jin03,FermionCondensation,MolcularProbe,JuhaCrossover,MolecularPop,DressedMolecules,Mackie05,Ohashi,ExactCrossover,CrossoverTemp,VorticesBECBCS,RotFermi}.

So far, most of the experimental studies involving cold atoms were
conducted in continuum. One of the major goals of studying
ultracold gases in optical lattices is to understand the physics
in condensed matter systems. In addition to topics already
investigated in continuum systems there are many effects of
interest lately: Superfluid to Mott insulator
transition~\cite{MottGreiner}, Bloch oscillation of particles in
lattice due to Bragg scattering~\cite{Kittel}, parametric atomic
down conversion in BEC ~\cite{ParametricLatt}, and so on.

In this letter, we will address the effect of a lattice on the
rotation of atoms in a harmonic trap. Many phenomena of rotation atoms in a continuous trap have already been investigated, both
experimentally and theoretically: the appearance of
vortices~\cite{VortexNucleation,FastRotBEC} in BEC and
BCS samples \cite{VorticesBECBCS}, quantum Hall
states for fast rotating fermions \cite{HoRotFermi}, and
vortex lattices in the lowest Landau level for
BEC~\cite{FastRotBEC,VortexLatticeLLL}. On the other hand,
lattices lead to many new effects under rotation, such as
structural phase transitions of vortex
matter~\cite{StructuralPhaseTransOL}. Also, near the
superfluid--Mott insulator transition, the vortex core has a
tendency toward the Mott insulating
phase~\cite{WuVortexconfigurations}, and second-order quantum phase
transitions between states of different symmetries were observed
at discrete rotation frequencies~\cite{CarrRot}.

In particular, it is well known that in a continuum the
centrifugal force prevents atoms from rotating beyond
the harmonic trap frequency. That means that if the rotation is too
fast the centrifugal force lets the atoms escape the
trap. Therefore, a quadratic-plus-quartic potential was assumed to
prevent the atoms from flying away from the trap at fast rotation
frequencies~\cite{RotQuatic}. In a lattice, however, as will be
shown in this paper, it is possible for atoms to stay in the trap
even if the rotation frequency is larger than the harmonic
trapping frequency and density depletion at the center of the trap
can then be developed for such a regime.

For completeness, we first review the rotation of a particle in a
two dimensional (2D) continuum. In the rotating frame, the
Hamiltonian for the particle is
\begin{equation} 
H_c=-\frac{\nabla^2}{2}+\frac{\omega^2\rho^2}{2}-\Omega L_z
\label{eq:ContinuumS}
\end{equation}
with $L_z=-i\partial/\partial_\phi$, $\omega$ is the trapping
frequency and $\Omega$ the rotation frequency. Throughout this
paper, units $\hbar=m=1$ are used, where $m$ is the
mass of the particle. Eq.~(\ref{eq:ContinuumS}) is formally
identical to the Hamiltonian of a particle of charge one placed in
a uniform magnetic field $2\Omega\hat{z}$ and confined in a
potential with a spring constant $\omega^2-\Omega^2$. This
equation can be solved by separation of variables ($r$ and $\phi$)
in polar coordinates where the
radial equation is given as 
\begin{equation} 
\left(-\frac{1}{2\rho}\frac{\partial}{\partial\rho}\rho\frac{\partial}{\partial\rho
}+\frac{\left(\omega^2-\Omega^2\right)\rho^2}{2}\right)\psi_r=\epsilon\psi_r.
\label{eq:ContinuumSradial}
\end{equation}
This decoupling of radial and angular degrees of freedom is different from the
motion in a lattice, which is to be discussed later. For
$\Omega\leq\omega$, Eq.~(\ref{eq:ContinuumS}) has eigenvalues
~\cite{FastRotBEC}
\begin{equation}
E_{j,k}=\omega+(\omega-\Omega)j+(\omega+\Omega)k
\label{eq:ConSpec}
\end{equation}
where $j,k$ are non-negative integers. Note that the angular
momentum states are eigenstates~\cite{FastRotBEC} such that there are
only level crossings when $\Omega$ changes. 
At $\Omega=\omega$, the ground states become infinitely degenerate lowest Landau levels. 

When $\Omega$ is bigger than $\omega$
($\omega^{\prime2}\equiv\Omega^2-\omega^2>0$), at
$r\rightarrow\infty$, the radial equation becomes
$\psi^{\prime\prime}_r+r^2\omega^{\prime2} \psi_r=0$, 
which gives the non-vanishing asymptotic solution
$\psi_r\propto\exp{[\pm i\omega^\prime r^2/2]}$. However, because
the harmonic trap potential at $r\rightarrow\infty$ approaches
infinity, the wavefunction at $r\rightarrow\infty$ has to vanish.
This contradiction indicates that at $\Omega>\omega$,
Eq.~(\ref{eq:ContinuumS}) has no solution. This means that the atoms would leave the trap because the centrifugal
force $\Omega^2r$ exceeds the restoring force $-\omega^2r$
in the xy plane.

\begin{figure}[t]
    \centerline{\includegraphics[clip,width=.8\linewidth]{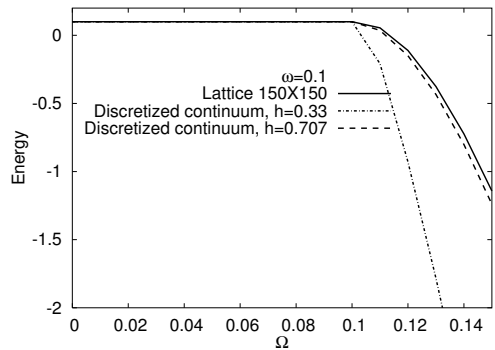}} 
    \caption{\protect\label{fig:ConVsLatTrap} Energy as function
of rotation frequency $\Omega$ in both the lattice and the
discretized continuum. The energy in the lattice is shifted by
$4t$, the energy at the bottom of the band. 
In a real continuum, the
atoms fly away from the trap at $\Omega
> \omega $ (see text). }
\end{figure}

\begin{figure}[ht]
\centerline{\includegraphics[clip,width=.8\linewidth]{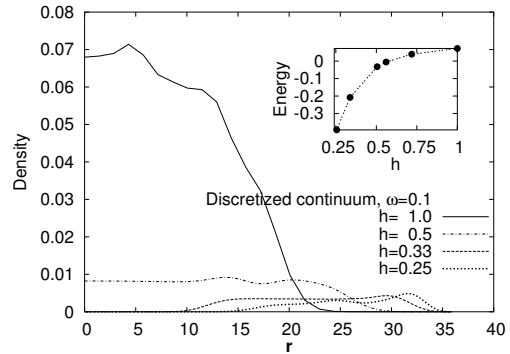} 
}
    \caption{\protect\label{fig:DiagDenConVsLat} Diagonal density
profile as function of the radius $r$ from the center of the trap
for the discretized continuum. When the rotation frequency
$\Omega$ is $0.11$, which is just above the trapping frequency
$\omega=0.1$, the finer the mesh size h, the closer the
wave function are pushed towards the boundary by the centrifugal
force. The inset, accordingly, shows that the energy drops when
the mesh gets
finer.} 
\end{figure}

Numerically, a system is treated by necessity as having a finite
size. Thus a procedure needs to be determined to distinguish a
``non-existing'' wave function (meaning that the particles have
left the trap) from a wave function where the particles are held
in the trap. To this end, we discretize the simple analytic
problem of Eq.~(\ref{eq:ContinuumS}) using a central difference
scheme with mesh size $h$. The results can be visualized like in
\fig{fig:ConVsLatTrap}: For well-contained particles the analytic
(\eq{eq:ConSpec}) and numeric (\fig{fig:ConVsLatTrap}) solutions
agree in that the ground state energy $E_g$ does not depend on the
rotation frequency $\Omega$. As soon as the particles are not
contained anymore (since the centrifugal force is now stronger than
the trap) the ground state energy plunges immediately. As can be
seen in \fig{fig:ConVsLatTrap}, this plunge happens slower for
larger mesh sizes. What is important here is that $E_g$ depends in
this (unphysical) case strongly on the mesh size. A similar case
can be made by looking at the probability density $|\psi_r|^2$  of
the particles in the trap, as shown in \fig{fig:DiagDenConVsLat}.
For the case of a rotation frequency too large to contain the
particles in the trap, it can be seen that only the fine-gridded
solution (here with the mesh size h$=0.25$) approaches our
physical understanding of the particles being driven outward by
putting the probability density all to the boundary of the
numerically available space. Therefore, the wavefunctions depend
on the mesh size only if the atoms are not contained and thus is
another good measure (along with a varying $E_g$) for this
situation.

We now study the case of zero-temperature atoms in a lattice
placed inside a harmonic trap. We will see that in this case, the
particles stay contained even
when the rotation frequency $\Omega$ is slightly larger than the
trapping frequency $\omega$. In the rotating frame, the single
band Hubbard Hamiltonian is~\cite{CarrRot}
\begin{equation}
\label{eq:LatticeH} H  =  \left[\sum_{\langle
i,j\rangle}\left(-t-i\Omega
K_{i,j}\right)c^+_ic_j+H.c.\right]+\sum_{i}V(\vec{r}_i)n_{i}
\end{equation}
where $\langle i,j\rangle$ indicates a sum over nearest neighbors,
$V(\vec{r}_i)=\Omega^2 r^2_i/2$ is the harmonic trap potential,
$t$ is the hopping term that describes tunneling between
neighboring sites, $n_i=c^+_ic_i$ is the number operator with
$c^+_i$ ($c_i$) the fermion creation (annihilation) operator at
site $i$, and $H.c.$ means Hermitian conjugate. In this paper, $t$
is set to one and used as a reference unit for $\omega$ and
$\Omega$. $\Omega K$ represents the centrifugal term with
$K_{i,j}=r_ir_j\sin\alpha_{i,j}/d^2$, where $r_i$ denotes the
distance from the axis of rotation to the $i$th site,
$\alpha_{i,j}$ is the angle subtended by the $i$th and $j$th sites
with respect to the axis of rotation, $d$ is the lattice constant,
and $\beta=0.493$ is the dimensionless constant characterizing the
lattice geometry and depth~\cite{CarrRot,Bhat06}. We note that,
different from that in continuum, the effective mass in the
lattice $\mu(k)=(2td^2\cos{kd})^{-1}$ depends on momentum $k$ and
may become negative. 

To avoid confusion here it should be noted that
without a trap, the particles will
immediately fly away independent of the magnitude of $\Omega$. This means that the lattice potential (in particular, the tunneling $t$ term) has absolutely no confining effect on the atoms. As a result, the addition of the lattice
potential to the trap potential cannot explain the
above-threshold rotation which will be explained here.

Figure \ref{fig:ConVsLatTrap} shows that in the presence of the
lattice $E_g$ does not depend on the rotation frequency for
$\Omega<\omega$ as it is the case for a continuum. For comparison,
in \fig{fig:ConVsLatTrap} we choose the lattice constant such that
$\mu(k=0)=1$ and find that the $\Omega$-dependence for
$\Omega>\omega$ of $E_g$ is the same as for the case of the
``discretized continuum'' (with the same effective mass $\mu=1$)
by setting the mesh size $h=1/\sqrt{2}$. The similarity in the
energy dependence for the lattice and discretized continuum let us
conclude 
that it is just a consequence of the
discrete nature of the lattice. Besides, it is to be remembered
that the infinite degeneracy of the ground state at
$\Omega=\omega$ in the continuum is lifted in the lattice and thus
the physics associated with the lowest Landau levels will be
different from that in the continuum. A detailed study of this
will follow in the future.

\begin{figure}[ht]
\vspace{-0.3in}
\centerline{\includegraphics[clip,width=.7\linewidth]{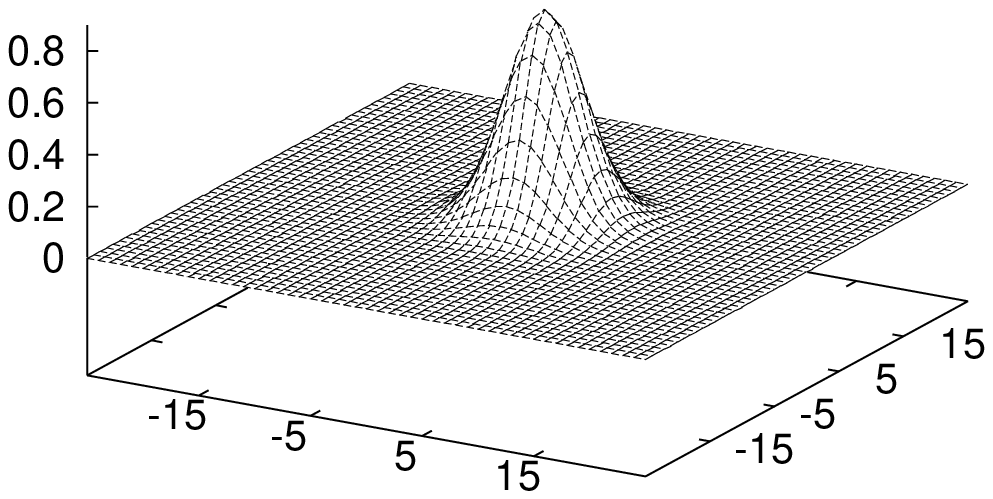}} 
\vspace{-0.65in}
\centerline{\includegraphics[clip,width=.7\linewidth]{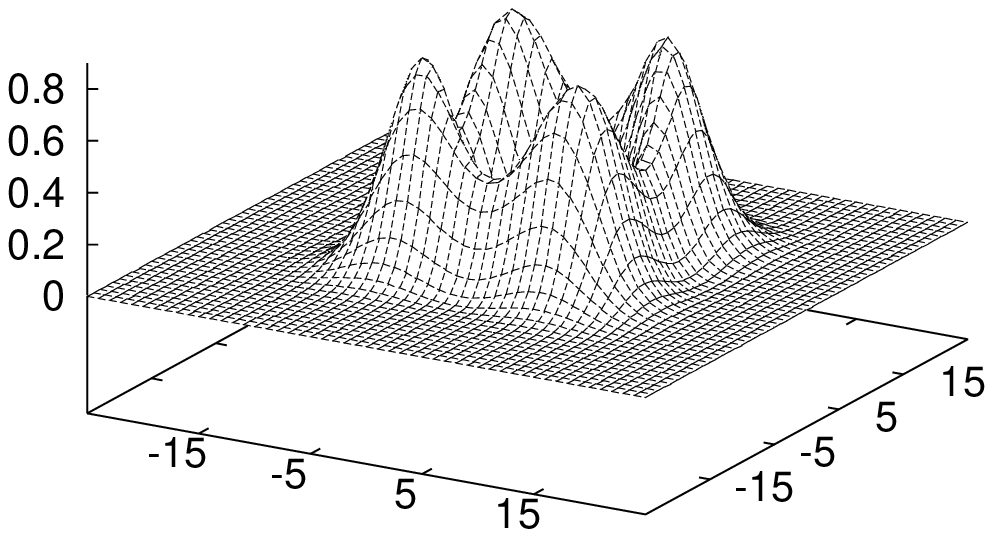}}
\vspace{-0.65in}
\centerline{\includegraphics[clip,width=.7\linewidth]{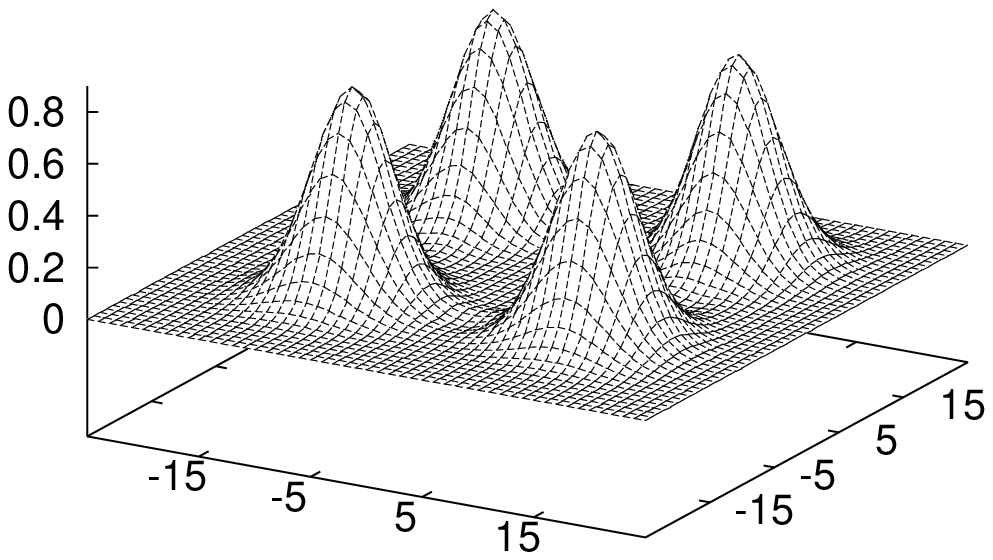} 
} \vspace{-0.3in}

    \caption{\protect\label{fig:LatticeThresholdBigTrap} Ground state
density distributions at $\Omega=0.09$ (a), $0.105$ (b) and $0.11$
(c). As the rotation frequency $\Omega$ becomes larger than the
trapping frequency $\omega=0.1$, the ground state wave functions
gradually change into the
4-lobe structure and the atoms are kept hold in the trap. 
For better visibility, the maximum density are
normalized to one for each plot. \vspace{-0.3in} }
\end{figure}


Figure~\ref{fig:LatticeThresholdBigTrap} shows the density
distribution at various $\Omega$ around $\omega=0.1$. Two features
deserve attention: First, the atoms still are
contained at $\Omega=0.11>\omega$ because if we increase the
lattice size, both the ground state wave function $|\psi_|^2$ and
 energies $E_g$ do not change.
At larger rotation frequency $\Omega$, different lattice
\myem{size/constants} lead to different density distributions and
ground state energies, which we interpret as particles escaping
the trap. In addition, the density at the center of the trap
becomes depleted, which is impossible for the ground state in the
continuum as shown in Ref.~\cite{HoRotFermi}. When the lowest four
states become degenerate, the ground state develops a 4-lobe
structure (Fig.~\ref{fig:LatticeThresholdBigTrap}c). Moreover,
when $\Omega$ increases beyond $\omega$, the ground states
continue to change smoothly and move away from the center
as shown in 
Fig.~\ref{fig:LatticeThresholdBigTrap}: 
the particles in the ground states only move away from the center
to a distance determined by the rotation frequency, but don't leave the trap altogether. It is also found that with the same $\omega$, the
particles in the ground state of a smaller lattice leave the trap at
smaller values of $\Omega$, as opposed to the universal threshold $\omega$
that is present in continuum. This difference between the lattice
and the continuum is one of the major results of this paper. Note
that the symmetric 4-lobe structure is determined by the primitive
cell geometry's fourfold symmetry, not by the overall lattice
shape, because, for example, 150X100 lattice gives the same
diagonal density distribution as 150X150 at $\omega=0.1$ and
$\Omega=0.11$. In other words, the fourfold degeneracy reflects
the discrete rotational symmetry of the underlying square
plaquette in the lattice~\cite{CarrRot}. Furthermore, when the
4-lobe structure is finally formed, both at $x=0$ and at $y=0$ the
densities approach zero.

\begin{figure}[ht]
\centerline{\includegraphics[clip,width=.7\linewidth]{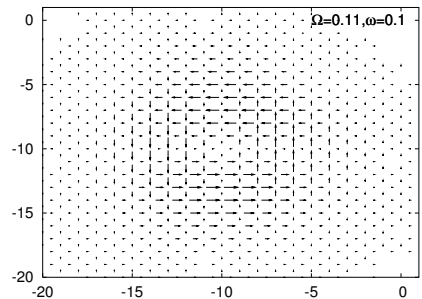} 
}
    \caption{\protect\label{fig:VecCurrentLobe} The current
distribution at one of the lobe in
Fig.~\ref{fig:LatticeThresholdBigTrap}c. A center for the current
pattern is clearly seen at site (-10,-10) by nearly disappearing
current at that site. However,
Fig.~\ref{fig:LatticeThresholdBigTrap}c shows that the density at
this site is nonzero, different from a vortex core. }
\end{figure}

The atoms rotating in the lattice are always in motion, so there
are always currents between neighboring sites, which are
calculated using $J_{ij}=i[n_i,H_{ij}]=it(a_ia^+_j-H.c.)+
\Omega(a_ia^+_j+H.c.)$~\cite{CarrRot}. One example of the current
pattern is shown in Fig.~\ref{fig:VecCurrentLobe}. While the
properties of the 4-lobe states will be for future publication,
the important thing here is that the persistent motion allows
Bragg scattering to explain the stabilization of atoms in the
lattice. 

To appreciate the role of Bragg scattering, we recall the Bloch
oscillations in the lattice~\cite{Kittel}: the driving voltage
across the lattice does not produce a net current for the
electrons, instead it produces periodic current, as the Bragg
scattering completely reflects the electrons. So it may not be so
surprising to see that the Bragg scattering can hold the atoms in
the trap even when $\Omega$ is slightly larger than $\omega$.
Furthermore, as the continuous symmetry is broken in the lattice,
the correlated motion in radial and angular direction allows Bragg
scattering between these directions. This is believed to also be the reason for the containment of the particles for rotation frequencies beyond $\Omega=\omega$. Additionally, we note that Bragg
scattering is not limited to the Hubbard Hamiltonian
Eq.~(\ref{eq:LatticeH}), because the discretized continuum
Hamiltonian Eq.~(\ref{eq:ContinuumS}) with large mesh sizes gives
similar results (the dashed line in Fig.~\ref{fig:ConVsLatTrap})
as that in lattice Eq.~(\ref{eq:LatticeH}).

\begin{figure}[ht]
    \centerline{\includegraphics[clip,width=.8\linewidth]{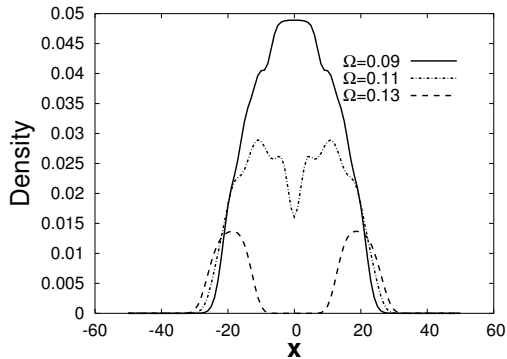} 
}
    \caption{\protect\label{fig:FermionsVortexTrap} Density
distribution of $50$ noninteracting spinless fermions at the cross
section with $y=0$. When $\Omega$ is close to but smaller than
$\omega=0.1$, the plateaus are developed reflecting the underlying
Landau-level wave functions~\cite{HoRotFermi}. On the other hand,
when $\Omega$ is larger than $\omega$, density at the center is
depleted. }
\end{figure}

So far, we have shown that the quantum statistics of atoms does
not play a role for the stabilization. Next, we point out that
density depletion at the center of the trap for spin polarized
fermions is possible. To this end, spin polarized fermions without
interaction are considered in the lattice at zero temperature. The
overall density of putting $N$ rotating fermions into a lattice,
according to Pauli's exclusion principle, is the summation of the
density of the lowest $N$ states. Note that the very low lying
states all have zero population at the center of the trap, which
causes the non-interacting fermions not only to have a density
plateau at $\Omega<\omega$ (the solid curve in
Fig.~\ref{fig:FermionsVortexTrap}) like  in the
continuum~\cite{HoRotFermi}, but one also sees density depletion
at $\Omega>\omega$ as shown by dashed line and dot-dashed line in
Fig.~\ref{fig:FermionsVortexTrap}. This is another major result of
this paper. In comparison, the density depletion cannot be
realized in a continuum with a harmonic trap because the atoms fly
away at $\Omega>\omega$.

In the above, coupling between the angular motion and radial
motion as well as Bragg scattering were used to explain the
stabilization. One may also think that the effective mass,
which depends on $k$ and could be negative in the lattice,
may also stabilize the atoms. 
However, the wave packets with $\mu(k)>0$ will not  be contained
in the trap at $\Omega>\omega$ if only the effective mass
contributes to the stabilization. Therefore, while it may play an
important role, the effective mass alone cannot explain the
stabilization.

In lattices, when $\omega$ is comparable to or greater than $t$,
the coupling between the angular motion and radial motion is
enhanced. As this enhancement makes the Bragg scattering between
radial direction to angular direction larger, the atoms will stay
in the lattice with larger rotation frequency. This means that to
experimentally demonstrate the above threshold rotation, larger
trapping frequency is desirable. However, deep harmonic trap makes
the single band Hubbard model not be reasonable for describing an
optical lattice system~\cite{Zoller98}. Therefore, some
intermediate trapping frequency is preferred for demonstrating the
above-threshold rotation experimentally.

To conclude, the rotation of atoms in an optical lattice is studied using a
Hubbard model. It is found that the atoms are contained in the
trap even when the rotation frequency exceeds the trapping
frequency, which is very different from the continuum case. Bragg
scattering and the coupling of angular and radial
motion make the above stability possible. In this regime, density
depletion at the center of the trap can be developed for spin
polarized fermions.



This work is supported by NSF and Research Cooperation. T. Wang
acknowledges the helpful discussions with J. Javanainen and U.
Shrestha.

\bibliography{BECBCS}

\end{document}